 \documentclass[prb,twocolumn,showpacs,amsmath,amssymb]{revtex4}
\def\LamF{{\lambda_{\rm F}}}

\newcommand{\eqbreak}{
\end{multicols}
\widetext
\noindent
\rule{.48\linewidth}{.1mm}\rule{.1mm}{.1cm}
}

\newcommand{\nf}{n_{\rm F}}
\newcommand{\deff}{\Delta}
\newcommand{\ha}{H_{\rm A}}

\newcommand{\curA}{{\cal A}}

\newcommand{\phdag}{{\phantom{\dagger}}}

\newcommand{\pf}{p_{\rm F}}

\newcommand{\Ef}{p_{\rm F}^{2}}

\newcommand{\rc}{{\bf r}_{\rm c}}

\newcommand{\imag}{{\rm Im}}

\newcommand{\curS}{{\cal S}}

\newcommand{\length}{l}
\input{epsf}
\usepackage{graphicx}
\usepackage{dcolumn}
\usepackage{bm}
\begin{document}
\title{Microwave conductivity of a d-wave superconductor disordered by extended impurities: a real-space
renormalization group approach}
\author{
Daniel E.~Sheehy }\email{sheehy@physics.ubc.ca}
\affiliation{
Department of Physics and Astronomy, 
University of British Columbia, 
6224 Agricultural Road, Vancouver, B.C.~V6T1Z1, Canada
}
  \date{March 11, 2003}

\begin{abstract}
Using a real-space renormalization group (RSRG) technique, the
microwave conductivity of a d-wave superconductor
disordered by extended impurities is calculated.
To do this, a semiclassical 
approximation is invoked which naturally accesses the Andreev bound
states localized near each impurity.
Tunneling corrections (which are captured using the RSRG)
lead to a delocalization of these quasiparticles
and an associated contribution to the microwave conductivity.

\end{abstract}
%
%
%
%
%
%
%
%
\maketitle
%
%

\section{Introduction}
\label{SEC:intro}
One aspect of the electronic properties of the cuprate superconductors which
remains mysterious is their frequency-dependent microwave 
conductivity~\cite{REF:Hosseini,REF:Turner}, the measurement of which sheds 
light on the effect of disorder on 
quasiparticle properties. 
 A common feature of many theoretical 
approaches to disorder (see, e.g., Refs.~\onlinecite{REF:Hirschfeld,REF:Durst})
in d-wave superconductors (although there have been a number of 
exceptions~\cite{REF:Balatsky,REF:Inanc,REF:Inanc2,REF:Durst2}) is the assumption that 
the scattering may be modelled by {\it pointlike} impurities.
Recent work~\cite{REF:Inanc2} found that the spectrum of quasiparticle excitations
is qualitatively different when the disorder is taken to consist of 
{\it extended\/} impurities, 
i.e., impurities characterized by
a typical size $a \gg \lambda_{\rm F}$, where  $\lambda_{\rm F}$ is the 
Fermi wavelength.  
 While the density of states $\rho(E)$ 
is believed to vanish linearly at low energies for the case of poinlike impurities~\cite{REF:Yashenkin01,REF:Altland}, in
Ref.~\onlinecite{REF:Inanc2} it  was found that, for disorder consisting of extended impurities, 
$\rho(E)$ is divergent ($\sim 1/E\log^3 E$)
at low-energies.
This low-energy buildup of states arises from the hybridization of Andreev bound states occurring
near each such extended impurity.  

Before proceeding, we provide two specific experimental motivations for studying
the effect of extended impurities on the electronic properties of d-wave superconductors:
Firstly, recent technical advances have allowed the fabrication of ultra-pure YBCO
 which has almost no atomic disorder in the CuO$_2$ 
planes~\cite{REF:Turner}.  A significant source of disorder 
 may originate away from the CuO$_2$ planes 
and can be expected to project a relatively 
long-wavelength potential on the CuO$_2$ planes.
Secondly, 
there is also the possibility of deliberately creating extended defects via 
ion irradiation techniques~\cite{REF:Walter}.
Apart from these specific possibilities, we note that although the calculation presented here 
strictly applies to the case of disorder satisfying $a \gg \lambda_{\rm F}$, the physical effects
may persist beyond this regime.

The purpose of this Paper is to further explore the consequences of extended impurities 
on the electronic properties of a d-wave superconductor by computing 
the associated contribution to the  microwave conductivity.
 To accomplish this, we invoke a semiclassical approximation~\cite{REF:Inanc,REF:Sheehy,REF:Inanc2} which accesses 
the low-energy Andreev-bound states 
occuring near each impurity.
Although the individual Andreev bound state wavefunctions are localized near the impurities, 
they become delocalized at low energies due to tunneling corrections~\cite{REF:Inanc2}, 
leading to a concomitant contribution to the conductivity.
We find that a natural way to study such tunneling effects between Andreev bound states 
(and to compute their associated contribution to the conductivity) is via a 
real-space renormalization group (RSRG) technique which is a variant of the decimation technique 
familiar from random spin systems~\cite{REF:Dasgupta,REF:Fisher,REF:Fabrizio,REF:Steiner,REF:Motrunich}.
Thus, although the Andreev bound states near individual impurities are strongly coupled to each other, the
RSRG reveals effective Andreev bound states associated with {\it many\/} impurities which are weakly
coupled and which contribute strongly to the low-energy microwave response.

This paper is organized as follows.  In Sec.~\ref{SEC:bdg}  we employ a semiclassical approximation which 
reduces the Bogoliubov-de Gennes eigenproblem for a disordered d-wave superconductor to a 
family of 
one dimensional
random pair potential  models (represented by a Hamiltonian $H_{\rm A}$).  In Sec.~\ref{SEC:rsrg} we introduce 
the RSRG decimation procedure for computing average properties of $H_{\rm A}$.  In Sec.~\ref{SEC:micro} we apply
this procedure to the computation of the microwave conductivity of a d-wave superconductor and discuss our 
results.  
In Sec.~\ref{SEC:concl}, we make some brief concluding remarks.
In the Appendix, we present a calculation of the average single-particle Green function
of $H_{\rm A}$.  

\section{Quasiparticles of a disordered d-wave superconductor}
\label{SEC:bdg}

We assume that the quasiparticles of a d-wave superconductor with extended impurities 
are governed by the following Bogoliubov-de Gennes (BdG) action:
\begin{equation}
{\curS} =  \int_0^{\beta} d\tau\int d^2 r  
\Psi^{\dagger} \left[ \partial_{\tau} +  \begin{pmatrix}
 \hat{h} &\hat{\Delta}   \cr
      \hat{\Delta} & -\hat{h}
\end{pmatrix}\right]\Psi,
\label{eq:action}
\end{equation}
where $\hat{h} \equiv -\nabla^2 - \epsilon_{\rm F}+ V({\bf r}) $, 
 $\epsilon_{\rm F} = \Ef$ is the Fermi energy
[i.e.~$\pf$ ($\equiv 2\pi/\LamF$) is the Fermi wavevector],
and we have adopted units in which $\hbar^2/2m=1$.
The d-wave pair potential operator $\hat{\Delta} =\Delta_0 (p_x^2- p_y^2)/\pf^2$ 
 with $\Delta_0$ being the 
pair-potential maximum. 
We take the disorder potential $V({\bf r})$ to arise from randomly located impurities of 
typical size $a$ and typical spacing $\ell$, and focus on the regime $\ell \gg a$.  

To study the Andreev bound states occuring near each extended impurity, 
we next invoke a semiclassical approximation (valid for $\pf a \gg 1$)  which has been extensively
discussed elsewhere~\cite{REF:Inanc,REF:Sheehy,REF:Inanc2} and which exchanges the BdG Hamiltonian for 
an effective problem residing on a classical trajectory
$\rc(s) =  (x_{\rm c}(s),y_{\rm c}(s))$ solving Newton's equation
 [i.e.~$2\Ef\,\partial_{s}^{2}\rc(s)= -{\bm{\nabla}}V(\rc(s))$]. 
This is 
implemented by writing for a particular incoming momentum direction $\hat{n} $ ($= \dot{\bf r}_{\rm c}(s)$
for $s\to -\infty$)
the field $\Psi$  as
\begin{equation}
\label{eq:eikonal}
\Psi = \curA {\rm e}^{i\pf S}\psi,
\end{equation}
and expanding to leading order in derivatives.  Here, 
the eikonal $S$  satisfies ${\bm{\nabla}}S = \dot{\bf r}_{\rm c}(s) $
with the overdot denoting differentiation with respect to the parameter $s$ along a trajectory.
The amplitude $\curA \approx 1$.  A given realization of $V({\bf r})$ leads to a 
set of classical trajectories labelled by $\hat{n}$ and an impact parameter $b$.

By inserting Eq.~(\ref{eq:eikonal}) into Eq.~(\ref{eq:action}), we see that 
the dynamics of the trajectory-dependent quasiparticles (represented by the field $\psi$) is governed by 
the Andreev~\cite{REF:Andreev} Hamiltonian $ H_{\rm A}$ 
\begin{equation}
\label{eq:andreev}H_{\rm A} =
\begin{pmatrix}
 -2i\pf\partial_s &  \deff(s) \cr
         \deff(s)  & 2i\pf\partial_s 
\end{pmatrix},
\end{equation}
which has the form of  a one-dimensional superconductor with an 
effective pair potential  
$\deff(s) = \Delta_0 ( \dot{x}_{\rm c}(s)^2 - \dot{y}_{\rm c}(s)^2)$.
For a given classical trajectory, $\deff(s)$ exhibits rapid variations that we 
wish to treat via an averaging procedure.  
To properly compute the disorder-averaged correlators of $\Delta(s)$ 
 one must solve Newton's equation for $\dot{\bf r}_{\rm c}(s)$ in the presence
of a given realization of $V({\bf r})$ and then average over all such realizations.  We
shall not attempt such a difficult calculation; however, on general grounds we expect that 
$\deff(s)$ has zero mean and short-range correlations beyond $\ell$.
We shall analyze  $H_{\rm A}$ by studying the 
Andreev bound states associated with sign changes in $\Delta(s)$ using the RSRG.  Before 
discussing how this works, let us briefly review previous results for this one-dimensional 
random pair-potential Hamiltonian (we note that $H_{\rm A}$ has appeared in several other 
condensed-matter contexts~\cite{REF:papers}).  
The  disorder-averaged low-energy 
density of states $\rho(E) \sim 1/E \ln^3 E$ 
of $H_{\rm A}$ was first obtained by Ovchinnikov and  \'Erikhman~\cite{REF:OE}
using a Fokker-Planck equation.  In Ref.~\onlinecite{REF:Inanc2}, this same result was obtained by 
mapping  $H_{\rm A}$ to a one-dimensional random hopping model.
More recently, the  disorder-averaged single-particle 
Green function  $\bar{G}(x-y;E)$ of $H_{\rm A}$  was obtained by Balents and Fisher~\cite{REF:BF} using a
supersymmetry method.  They found that  $\bar{G}(x-y;E)  \sim \exp[-|x-y|/\log^2 E]$ for 
$|x-y| \to \infty$. In the Appendix, we reproduce these results using the 
RSRG technique; here we note that, in the present context, this behavior signifies that the Andreev
bound state wavefunctions are  exponentially localized but with a characteristic length scale that 
{\it diverges\/} at low energies (i.e.~they become delocalized).  
Finally, as discussed in Refs~\onlinecite{REF:Inanc,REF:Inanc2},
  $H_{\rm A}$ is closely related to models of  supersymmetric quantum mechanics~\cite{REF:Witten,REF:SvH82}.

\section{Real-space Renormalization group approach}
\label{SEC:rsrg}

\begin{figure}[hbt]
%
 \epsfxsize=\columnwidth
%
\centerline{\epsfbox{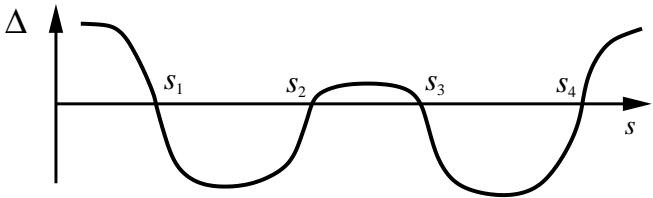}}
\vskip0.50cm \caption{ Sketch of the effective pair potential
along a trajectory.  The pair-potential has been deliberately drawn smaller between 
sites $s_2$ and $s_3$ to illustrate a point; see the text for details. }
\label{fig1}
\end{figure}

In this section we discuss how the disorder-averaged low-energy properties of $H_{\rm A}$ may be analyzed via 
a RSRG approach.  Of central importance in such an analysis are the sign-changes (i.e., zeroes) of 
$\Delta(s)$~\cite{note:zero}.
Henceforth, our qualitative picture of  $\Delta(s)$ along a trajectory is depicted in Fig.~\ref{fig1}, i.e.,
it exhibits many sign changes.
Before considering the many sign-change case, however, let us consider an infinite system in which $\Delta$ has exactly one 
sign change.  In this case, 
 $\ha$ has an exact zero-energy (ZE) eigenstate of the form~\cite{REF:Witten,REF:SvH82,REF:Inanc}
\begin{equation}
\Psi_n(s) \propto   \begin{pmatrix}
1 \cr \pm i\end{pmatrix}
{\rm e}^{\pm \frac{1}{2\pf} \int_{s_n}^s \deff(s)} ,
\label{eq:susystate}
\end{equation}
where the $+(-)$ corresponds to the case $\deff(s) < 0 $  ($\deff(s) > 0 $)
for $s>s_n$ and 
we have omitted an overall normalization factor.  
Indeed, the normalizability of Eq.~(\ref{eq:susystate}) relies on the existence 
of the sign change in $\Delta(s)$.
 The subscript $n$ enters when we consider the case in which the pair potential has 
multiple sign changes; henceforth it 
shall refer to the $n$th zero $s_n$ of  $\Delta(s)$ and refer to $\Psi_n(s)$ as the  ZE 
state associated with $s_n$.
In the case in which $\Delta(s)$ has multiple sign changes, however, we must account for 
tunneling corrections $t_n = \langle \Psi_n | \ha | \Psi_{n+1}  \rangle $ between nearest-neighbor 
ZE states (i.e.~they are no longer at zero energy).  A direct calculation gives~\cite{REF:Inanc}
\begin{eqnarray}
\label{eq:tee}
&&t_n =   \sqrt{\frac{2\pf}{\pi}} 
|\deff'(s_n) \deff'(s_{n+1})|^{1/4} \\
&&\qquad \times
\exp\left(- \frac{1}{2\pf} \left| \int^{s_{n+1}}_{s_n} \deff(s)ds
\right|\right) \nonumber,
\end{eqnarray}
where the prime indicates differentiation with respect to $s$. 

Having identified the ZE states $\Psi_n$ associated with the zeroes of $\Delta(s)$, 
we now turn to the RSRG analysis of such states. Although it is formally similar to that studied in 
Refs.~\onlinecite{REF:Dasgupta,REF:Fisher,REF:Fabrizio,REF:Steiner,REF:Motrunich}, there are conceptual differences
which are important. The essential result of such an analysis is
the following: Although the matrix elements $t_n$ coupling the $\Psi_n$ may be large, there 
are effective ZE states associated with {\it many\/} sign changes which are weakly coupled and 
which may be treated via perturbation theory.  Furthermore, the couplings between 
such effective ZE states have a probability distribution which is universal in a sense discussed below.
To begin, we note that
for a given realization of the pair potential, there exists a strongest matrix element $t_{\rm max}$. 
The ZE states associated with
this matrix element are strongly coupled, forming symmetric and antisymmetric wavefunctions at energies $\pm t_{\rm max}$. 
The difference between these energies defines our initial bandwidth $\Omega_0 = 2  t_{\rm max}$.  To 
estimate the magnitude of $\Omega_0$, we first assume the impurities are dense enough such that the exponential 
factor in Eq.~(\ref{eq:tee}) is negligible.  Then, we assume that $\Delta'(s_n)\approx \Delta_0/a$ near an extended
impurity of size $a$.  Using $\Delta_0\sim 50 meV$, we have $\Omega_0 \sim 10^{14} s^{-1} /\sqrt{\pf a}$.

Generally, we are interested
in an observable at some low-energy $\omega \ll \Omega_0$; the RSRG progressively eliminates all pairs of ZE
states coupled by matrix elements larger than $\omega$. 
To see what this means, let us examine Fig.~\ref{fig1}.  According to Eq.~(\ref{eq:tee}), 
the matrix elements $t_n$ are exponentially dependent on the pair-potential integral 
$|\int_{s_n}^{s_{n+1}} \Delta(s) ds|$  between pairs of sites.    Thus, in Fig.~\ref{fig1}, the strongest 
matrix element is 
that connecting sites $s_2$ and $s_3$, (i.e., $t_2$) since the aforementioned integral is smallest for them.
To obtain the decimation procedure, we recall that the 
the sites $s_n$ are {\it defined}  by the sign changes of $\deff(s)$ along a particular trajectory.  The 
fact that $\deff$ is only slightly above zero for the segment between $s_2$ and $s_3$ suggests
that
we could have simply {\it ignored\/} the sign changes at  $s_2$ and $s_3$ without making 
any appreciable error in computing physical quantities.  This defines a new description of $H_{\rm A}$:
 a basis of 
ZE states that are associated with all sign changes of $\Delta(s)$ {\it except\/} 
those at sites $s_2$ and $s_3$.  The effect of these zeroes of $\Delta(s)$  
is incorporated into a weak matrix element between the 
{\it effective\/} ZE states at sites $s_1$ and $s_4$  (that are each associated with 
three sign changes of $\Delta(s)$).   
This procedure exchanges
four   ZE states for two weakly-coupled ZE states 
(and lowers the bandwidth to  $\Omega < \Omega_0$).  The associated matrix element 
$\tilde{t}$ between the effective ZE states at sites $s_1$ and $s_4$ is still given 
by the general formula Eq.~(\ref{eq:tee}); it is easy to see that this formula
has the following recursion property:
\begin{equation}
\label{eq:recursion}
\tilde{t} = t_1 t_3 /t_2.
\end{equation}

Remarkably, this recursion relation is the same as that obtained by Fisher~\cite{REF:Fisher}
for the random antiferromagnetic spin chain; the existence of such a 
relation is the reason the RSRG works in this seemingly unrelated system.  
 It is convenient to define 
(as was done in the random spin chain case~\cite{REF:Fisher}) logarithmic matrix-element strengths 
$\zeta_n \equiv \log (\Omega/2) - \log t_n$
and arc lengths $\length_n \equiv s_{n+1}- s_n$.
It is useful to note that 
up to a constant (defined so that the minimum $\zeta_n$ is zero), $\zeta_n$ is essentially the abovementioned 
pair-potential integral $|\int_{s_n}^{s_{n+1}} \Delta(s) ds|$ that motivated the decimation scheme. 
%
%
%
A particular (trajectory-dependent) form for $\ha$ defines an initial probability distribution 
$P(\zeta,\length;\Omega_0)$
for the $\{\zeta_n\}$ and $\{ \length_n \}$, in which ZE states have been  assigned to all sign changes of $\Delta$.
How do these values evolve under the RSRG decimation?  Once sites $s_2$ and 
$s_3$ have been decimated, the new effective arc length is clearly  $\tilde{\length} = s_4 - s_1$.  This, along 
with Eq.~(\ref{eq:recursion}), implies that
\begin{subequations}
\begin{eqnarray}
\label{eq:recur1}
\tilde{\zeta}&=& \zeta_1+\zeta_3-\zeta_2 = \zeta_1+\zeta_3,
\\
\tilde{\length}&=& \length_1+\length_2+\length_3,
\label{eq:recur2}
\end{eqnarray}
\end{subequations}
where in Eq.~(\ref{eq:recur1}) we have used the fact that (as noted in Ref.~\onlinecite{REF:Fisher})
 since the matrix element connecting $s_2$ and
$s_3$ is assumed to be the strongest, $\zeta_2 \equiv 0$ by definition.
By iterating this procedure, we produce a basis set of ZE states 
 which are more weakly coupled 
each iteration (and which are associated with progressively more sign changes of
$\Delta(s)$) but which still contain all the low energy physics of $H_{\rm A}$.
 As mentioned above, although their physical origin is slightly different, the recursion rules for 
the $\{ \zeta_n \}$ and the  $\{ \length_n \}$ are the same as those studied by 
Fisher for random spin chains.  
The essential results of this work (for our purposes) 
are the number of sites  $n_{\Omega} = 1/ \ln^2 \Omega_0/\Omega$
remaining after having lowered the bandwidth (by decimating) to 
energy $\Omega$ and the asymptotic ($\Omega \to 0$) probability distribution $P(\zeta,\length;\Omega)$,
 which is given by~\cite{REF:Fisher}
\begin{subequations}
\begin{eqnarray}
\label{eq:dist}
P(\zeta,\length;\Omega) &=& \frac{1}{\Gamma^3} Q(\zeta/\Gamma,\length/\Gamma^2),
\\
\hat{Q}(\eta,\hat{y})& =& \frac{\sqrt{c\hat{y}}}{\sinh{\sqrt{c\hat{y}}}} 
{\rm e}^{-\eta\sqrt{c\hat{y}}\coth{\sqrt{c\hat{y}}}}.
\end{eqnarray}
\end{subequations}
Here, $\hat{Q}(\eta,\hat{y})$ is the Laplace transform of $Q(\eta,y)$
and $\Gamma \equiv \ln \Omega_0/\Omega$.  The integration constant $c$
defines the length scale over which $\length$ varies.  We
assume this is given by the typical value of $\length$ in the initial distribution, i.e., 
$c\approx \ell$.

\section{microwave conductivity}
\label{SEC:micro}

Having outlined the RSRG technique, we now apply it to the calcuation of a specific quantity:
The microwave conductivity  of a d-wave superconductor disordered by extended impurities, 
averaged over all realizations of the disorder.
Before embarking on the calculation, let us discuss the physical picture.  We are interested in the possibility
of the Andreev bound state quasiparticles contributing to the microwave conductivity.  From the outset
it is not obvious that they will do so, 
since they are nominally localized [see Eq.~(\ref{eq:susystate})] near a particular sign change. 
The results of Balents and Fisher for the Green function $\bar{G}(x-y;E)$ of $H_{\rm A}$  give hope to this possibility, 
revealing that the quasiparticles become delocalized at asymptotically low energies.
As the conductivity is related to a two-particle Green function (see below), the results of Ref.~\onlinecite{REF:BF}
are not sufficient to compute the conductivity.
However, as we shall see
the peculiar way in which tunneling corrections conspire to form {\it effective\/} Andreev bound states
associated with many sign changes conspires to give an interesting contribution to the 
microwave conductivity.

For a given realization of the disorder, the real part of the microwave 
conductivity $\sigma_1$ is given by the Kubo formula~\cite{REF:Mahan}:
\begin{subequations}
\begin{eqnarray}
\sigma_1(\omega) &=&- \frac{1}{\omega} \imag\Pi^{\rm R}(\omega),
\\
\label{eq:pi}
\Pi(i\Omega) &\equiv&
 - \frac{1}{A} 
\int_0^{\beta} d\tau {\rm e}^{i\Omega \tau} 
\langle {\bf j}({\bf q},\tau) \cdot
{\bf j}({\bf q},0) \rangle|_{{\bf q}\to 0},
\end{eqnarray}
\end{subequations}
   Here,
$A$ is the area of the system and to obtain the retarded polarization tensor we use the 
usual analytic continuation
 $\Pi^{\rm R}(\omega) =  \Pi(i\Omega \to \omega + i0^+)$.  The angle brackets 
represent the thermodynamic average with respect to the action $\curS$. We shall denote the disorder
average by square brackets; however, before attempting to evaluate this average we first 
simplify Eq.~(\ref{eq:pi}) for $\Pi(i\Omega)$.

Within the semiclassical approximation of Sec.~\ref{SEC:bdg},  ${\bf j}({\bf q},\tau)$ 
is obtained by inserting Eq.~(\ref{eq:eikonal}) into the usual 
current operator ${\bf j}({\bf r},\tau) = ie(\Psi^{\dagger}\bm{\nabla} \Psi - h.c.)$.  The  
Fourier transform ${\bf j}({\bf q},\tau)$ is  
(in the ${\bf q} \to {\bf 0}$ limit)
\begin{equation}
\label{eq:current}
{\bf j}({\bf 0},\tau)  = -e\pf^2 \sum_{\hat{n}}\int db \,ds\,  \dot{\bf r}_{\rm c}(s)\,
\psi^{\dagger}(s,\tau) \psi(s,\tau).
\end{equation}
Standard manipulations then yield the following expression for $\sigma_1 (\omega)$:
\begin{subequations}
\begin{eqnarray}
&&\sigma_1(\omega) =  
 \frac{e^2 k_{\rm F}^3}{\omega A} \sum_{\hat{n}}
\int db  ds  ds' \dot{\bf r}_{\rm c}(s) \cdot  \dot{\bf r}_{\rm c}(s') \imag K^{\rm R}(s,s';\omega),
\nonumber
\\ 
\label{eq:sigmafin}
\\
&&K(s,s';\tau) \equiv \langle \psi^{\dagger}(s,\tau) \psi(s,\tau) 
 \psi^{\dagger}(s',0) \psi(s',0) \rangle,
\label{eq:pi2}
\end{eqnarray}
\end{subequations}
   Next, we use Wick's theorem ($\curS$ is a Gaussian action) to express the Matsubara
Fourier transform $K(s,s';i \Omega)$ of Eq.~(\ref{eq:pi2})
as a sum of two terms, one of which gives a vanishing contribution to $\Pi$.  The other contribution
to $K(s,s';i \Omega)$ may conveniently be expressed in terms of the  normalized 
eigenstates $\{\psi_i(s)\}$ (with eigenvalues $\{ E_i \}$) of $H_{\rm A}$:
\begin{eqnarray}
&&\imag K^{\rm R}(s,s';\omega) = \pi\sum_{i,j}  \nonumber 
\psi^{\dagger}_i(s) \psi^{\phdag}_j(s) \psi^{\dagger}_j(s') \psi^{\phdag}_i(s')
\\
 &&\qquad 
\times(\nf(E_i) - \nf(E_j))
\delta(\omega + E_i - E_j) ,
\label{eq:imagk}
\end{eqnarray}
where, since we are interested in $\imag\, \Pi^{\rm R} (\omega)$, we have displayed  
$\imag\,  K^{\rm R}(s,s';\omega)$ directly.

Turning to Eq.~(\ref{eq:sigmafin}), we see that for the conductivity we require the disorder average of 
$\imag K^{\rm R}(s,s'; \omega)$ {\it multiplied\/} by $\dot{\bf r}_{\rm c}(s) \cdot  \dot{\bf r}_{\rm c}(s')$.  
At low energies, the ZE states hybridize and become delocalized along the classical
trajectories.  By contrast, the average of $\dot{\bf r}_{\rm c}(s) \cdot  \dot{\bf r}_{\rm c}(s')$ is expected
 to vanish on long length scales as the classical trajectories bounce off the extended impurities.  
Thus, it is sensible to approximate the average of the product of these quantities
by the product of the averages; furthermore we shall approximate for the classical problem
$[\dot{\bf r}_{\rm c}(s) \cdot  \dot{\bf r}_{\rm c}(s')]_{\rm dis.} \simeq \exp(-|s-s'|/\ell)$ i.e., 
exponentially decaying correlations.   We believe that the precise form of this correlator is unimportant 
at low energies. However, we note that the virtue of this form is that it naturally interpolates
between ballistic motion on short length scales 
(i.e.~$[({\bf r}_{\rm c}(s) - {\bf r}_{\rm c}(0))^2]_{\rm dis.} = s^2$ for $s \to 0$)
and diffusive motion on long length scales 
(i.e.~$[({\bf r}_{\rm c}(s) - {\bf r}_{\rm c}(0))^2]_{\rm dis.} = 2s/\ell$  for $s \to \infty$).

Next, we turn to the computation of the disorder average of 
$\imag K^{\rm R}(s,s'; \omega)$, for which we require the RSRG analysis.
The purpose of the RSRG is to eliminate (decimating until $\Omega = \omega$)
all the strong matrix elements so that we are left 
with an effective theory having only weak (i.e.~$t_n \alt \omega$) matrix elements between ZE states, 
allowing the use of perturbation theory.  Following Fisher~\cite{REF:Fisher}, we assume that 
the broadness of  $P(\zeta,l;\Omega)$ indictates that the disorder average 
is dominated by a class of rare pair-potential configurations that provide a large contribution.
For the disorder average of Eq.~(\ref{eq:imagk}),
the class we have in mind is when an undecimated pair of states is separated by  $\length = |s-s'|$
{\it exactly\/}; 
the relative number of such configurations with logarithmic matrix element 
strength $\zeta$ is simply given by $ n_{\omega}P(\zeta,|s-s'|;\omega)$.
 For a given $\zeta$ and to leading-order 
in perturbation theory 
 (a similar calculation was done in 
Ref.~\onlinecite{REF:Motrunich}), the ZE states at $s$ and $s'$ form symmetric ($S$) and 
antisymmetric ($A$) wavefunctions with energies $\pm t = \pm \frac{\omega}{2}{\rm e}^{-\zeta}$ (recall 
we have set $\Omega = \omega$).
At low energies, 
the only terms which contribute appreciably to the sum in Eq.~(\ref{eq:imagk}) are these states 
(i.e.~$i,j =S,A$). Thus, we have (upon summing over all such configurations which amounts to 
integrating over $\zeta$)
%
%
\begin{subequations}
\begin{eqnarray}
\label{eq:kfin}
\!\!\!\![\imag K^{\rm R}(s,s';\omega)]_{\rm dis.}\!  &\propto&\!
n_\omega \int_0^{\infty} d\zeta P(\zeta,|s-s'|;\omega)
\\
&&\times\delta(\omega - \omega{\rm e}^{-\zeta} )\tanh(\omega/4T), \nonumber
\\
&\propto&\!\! P(0,|s-s'|;\omega) 
\frac{\tanh(\omega/4T)}{\omega \log^2 \Omega_0/\omega} ,\label{eq:kfin2}
\end{eqnarray}
\end{subequations}
where we have omitted overall 
temperature- and frequency-independent prefactors, since the RSRG is not 
expected to capture these correctly.  In Eq.~(\ref{eq:kfin}), we see that
the delta function constraint restricts attention to $\zeta = 0$, i.e., 
to those pairs of states that are about to be decimated at energy $\omega$,
leading to Eq.~(\ref{eq:kfin2}).

Having obtained an expression for $[\imag K^{\rm R}(s,s';\omega)]_{\rm dis.} $, we now insert it into
Eq.~(\ref{eq:sigmafin}).  We can immediately evaluate one of the integrations over the parameters
$s$ and $s'$ giving the Laplace transform of $P(0,|s-s'|;\omega)$.  
Up to a constant of order unity, the 
remaining integrations $\int db ds \propto A$, so that we have 
\begin{equation}
\label{eq:final}
[\sigma_1(\omega)]_{\rm dis.} \simeq \sigma_0  \frac{\Omega_0}{\omega}
\frac{\tanh(\omega/4T)}{\log^2 \Omega_0/\omega},
\end{equation}
where $\sigma_0$ is a dimensionful prefactor that is difficult to estimate (in part due to the uncertainty
in the prefactor of $n_{\Omega}$); a very rough estimate along the lines of our estimate of $\Omega_0$
gives $\sigma_0 \sim e^2\epsilon_{\rm F}\Delta_0 \ell/\Omega_0^2 a$.
%

\begin{figure}[hbt]
%
 \epsfxsize=6.7cm
%
\centerline{\epsfbox{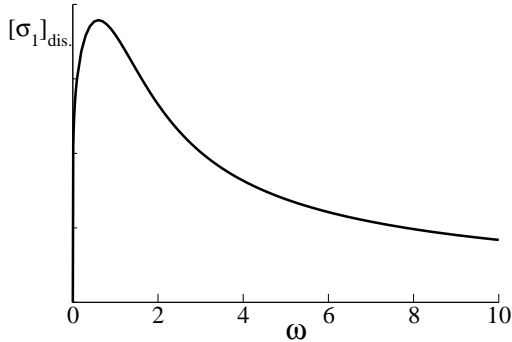}}
\vskip0.50cm \caption{ Plot of $[\sigma_1(\omega)]_{\rm dis.}$(in arbitrary units).  
Here, $\omega$ is measured in units of $10^9 s^{-1}$ and we have chosen  $T = 10 mK$ and 
$\Omega_0 = 10^{13}s^{-1} $.}
\label{fig2}
\end{figure}

This formula (plotted in Fig.~\ref{fig2}) applies at $\omega\ll \Omega_0$ and includes {\it only} the 
contribution due to extended impurities.   It exhibits a peak in $\omega$ which becomes larger and narrower and 
moves to zero as $T\to 0$.  Also, $ [\sigma_1(0)]_{\rm dis.}= 0 $. 
This contribution would occur in parallel to those arising from pointlike 
impurities~\cite{REF:Hirschfeld,REF:Durst}, and would therefore
be most noticeable at low temperatures where the peak is sharpest. 
For example, according to the calculations of Durst and Lee~\cite{REF:Durst}, 
the microwave conductivity due to pointlike impurities
should attain a universal value at low temperatures and frequencies, in striking contrast to the
structure of Eq.~(\ref{eq:final}) which would exhibit strong variations in this regime.
 At the typical temperature
scales of the Turner et al experiments~\cite{REF:Turner}, however, the extended impurity contribution would 
be particularly broad and small making a quantitative comparison difficult.   To isolate 
the contribution considered here, we make two suggestions: (1) This contribution 
is largest at $\omega \to 0$ with $T\alt \omega$ so that the argument of the $\tanh$ in 
Eq.~(\ref{eq:final}) is large. 
(2) By deliberately introducting extended impurities via ion 
irradiation~\cite{REF:Walter} or by perhaps introducing disorder away from the CuO$_2$ planes,
this contribution will be enhanced.
  
\section{Concluding remarks}
\label{SEC:concl}

In an effort to understand the electronic properties of disordered d-wave superconductors, we have computed
the contribution to the microwave conductivity arising from extended impurities.  This contribution has its origin in 
the hybridization of ZE Andreev bound states occurring near each such extended impurity.  
To capture this hybridization, we have applied a real space renormalization group technique that generalizes
the work of Refs.~\onlinecite{REF:Dasgupta,REF:Fisher,REF:Fabrizio,REF:Steiner,REF:Motrunich} on lattice 
models to the continuum Hamiltonian $H_{\rm A}$.  
From a theoretical point of view, this generalization is 
possible because the underlying supersymmetry of $H_{\rm A}$~\cite{REF:Inanc,REF:Witten,REF:SvH82} implies
that there is an approximate zero-energy state near {\it each\/} zero of $\Delta(s)$ for every realization of the
random pair potential, providing a natural basis for studying the low-energy properties of $H_{\rm A}$ via 
such tunneling corrections.
We found that the singular density of states associated with these Andreev bound states found in 
Ref.~\onlinecite{REF:Inanc2} leads to a singular contribution to the microwave conductivity and 
discussed the possiblity of observing this experimentally.
Finally, we note that whereas the low-energy quasiparticles studied here rely on the d-wave symmetry
 of the pair potential, they are not associated with the nodes of the pair potential.

\smallskip
\noindent
{\it Acknowledgments\/} ---  
We gratefully 
acknowledge useful discussions with L. Balents, D. Bonn, T. Davis, M. Franz, R. Harris, T. Pereg-Barnea,
S. Plotkin, P. Turner and K. van der Beek.
This work was supported by NSERC.

\appendix
\section{Single-particle Green function}

In the present section, we apply the RSRG technique to the calculation of the Green function
of $H_{\rm A}$ averaged over all realizations of the random pair potential. These results apply more generally
 to any  system in which  $H_{\rm A}$ arises~\cite{REF:papers,REF:OE,REF:BF}.
This calculation 
is similar to the one presented in the main body of the 
text and obtains a result which has already been obtained by
Balents and Fisher~\cite{REF:BF}.  Nevertheless we present it here for the sake of clarity and completeness, as well
as to highlight the strengths and weaknesses of the RSRG method. 
Indeed, the 
fact that we are able to reproduce (up to numerical prefactors) 
the exact results of Ref.~\onlinecite{REF:BF} lends support for this technique.  The 
disorder-averaged Green function  $\bar{G}(x-y;i\omega)$ is given by 
\begin{eqnarray}
\label{eq:green1}
&&\bar{G}(x-y;i\omega) = \left[ G(x,y;i\omega) \right]_{\rm dis.},
\\
&&(i\omega - H_{\rm A}) G(x,y;i\omega)  = \delta(x-y),
\end{eqnarray}
where the square brackets in Eq.~(\ref{eq:green1}) denote the average over
all realizations of the pair potential.  It is useful to express $G(x,y;i\omega)$ 
in terms of the normalized eigenstates  $\psi_n$ of 
 $H_{\rm A}$:
\begin{equation}
 G(x,y;i\omega) = \sum_n  \frac{\psi_n(x) \psi_n(y) }{i\omega - E_n}.
\label{eq:sum}
\end{equation}
Before proceeding with the RSRG calculation, we emphasize
that it does not correctly
obtain overall prefactors associated  with the contributions to $\bar{G}(x-y;i\omega)$.
Nonetheless, it obtains in a simple way the correct (nontrivial) asymptotic behavior of $\bar{G}(x-y;i\omega)$ as
well as the correct matrix structure.  In the following, these are the features we are primarily interested in 
obtaining.

The RSRG simplifies the computation of $\bar{G}(x-y;i\omega)$ in two distinct ways.  Firstly, by decimating 
until $\Omega = \omega$, we eliminate all strongly coupled  zero-energy (ZE) states. Since the remaining ZE 
states are weakly coupled, it is plausible to approximate the exact low-energy eigenstates 
$\psi_n(x)$ by perturbed ZE states.  Secondly, due to the fact that $P(\zeta,l;\omega)$ is very broad~\cite{REF:Fisher}, 
certain rare 
 pair-potential configurations 
dominate the average over all random pair-potential configurations.  Here, we focus on two leading
contributions; analagous contributions 
were considered in Ref.~\onlinecite{REF:Steiner}
in the context of the related problem of computing the single particle Green function for a one-dimensional 
random hopping problem. However, we obtain slightly different results for the second contribution.
 The first we denote by  $\bar{G}_1(x-y;i\omega)$.  This conribution
 is due to pair potential configurations such that 
 the  coordinates $x$ and $y$ are near sign changes that are about to be decimated  (owing
to the fact that the pair potential connecting them is relatively small) in the sense discussed 
in Sec.~\ref{SEC:rsrg}.  The second contribution,  $\bar{G}_2(x-y;i\omega)$, is due to pair potential
configurations having only one of  $x$ or $y$ be about to be decimated along with a ZE state at 
a third site.  Let us first, however, 
discuss  $\bar{G}_1(x-y;i\omega)$.

\begin{figure}[hbt]
%
 \epsfxsize=\columnwidth
%
\centerline{\epsfbox{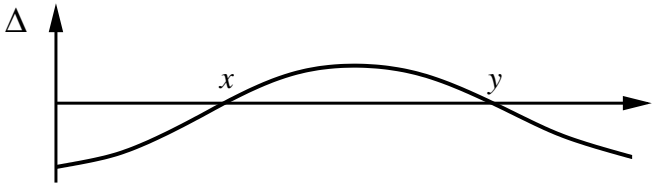}}
\vskip0.50cm \caption{Class of pair potential configurations providing the contribution
to $\bar{G}_1(x-y;i\omega)$. The smallness of $\Delta$ between $x$ and $y$ is meant to indicate 
that the ZE states associated with these sites are about to be decimated. }
\label{fig3}
\end{figure}

The rare pair-potential configuration that leads to $\bar{G}_1(x-y;i\omega)$ is the same 
as was considered in Sec.~\ref{SEC:micro} for 
$[\imag K^{\rm R}(s;\omega)]_{\rm dis.}$ and is 
depicted 
schematically in Fig.~\ref{fig3}; the fact that both $x$ and $y$ are about to be decimated at energy $\omega$ is 
depicted pictorially by having $\Delta$ be relatively low between $x$ and $y$.  Of course, it is not
necessary that the sites $x$ and $y$ be nearest-neighbor sign changes as depicted in Fig.~\ref{fig3}, but merely
that they are nearest neighbors  after having decimated until $\omega$.  The relative 
number of such 
configurations is given by $n_{\omega} P(0,|x-y|;\omega)$; to obtain $\bar{G}_1(x-y;i\omega)$ we  simply multiply
this by the magnitude of the contribution
given by Eq.~(\ref{eq:sum}).  
For concreteness, let us assume that $\Delta>0$ on average  (as in Fig.~\ref{fig3})
between $x$ and $y$ ($y>x$); once we have computed this it is straightforward to obtain the  $\Delta<0$ configurations.
  The ZE states associated with these points are given by 
Eq.~(\ref{eq:susystate})  with the $-$ sign for the ZE state $\Psi_x$ at $x$ and the $+$ sign for the 
ZE state $\Psi_y$ at $y$:
\begin{eqnarray}
\label{eq:spinor}
\Psi_{x} \propto  \begin{pmatrix}
1 \cr - i\end{pmatrix} {\rm e}^{i \phi_x}
\\
\Psi_{y} \propto  \begin{pmatrix}
1 \cr  i\end{pmatrix} {\rm e}^{i \phi_y},
\label{eq:spinor2}
\end{eqnarray}
where we have suppressed unimportant normalization factors and displayed phase factors in these expressions 
which were suppressed in Eq.~(\ref{eq:susystate}). These factors are chosen to 
satisfy ${\exp} i (\phi_x - \phi_y ) = -i$ so that $t_n$ in Eq.~(\ref{eq:tee}) is real and positive;
one must account for these properly to obtain the correct phase of $\bar{G}_1(x-y;i\omega)$.    
  Under the influence of the tunneling matrix element $t$, (which is small 
by virtue of the decimation procedure), these states form symmetric ($S$)  and antisymmetric  ($A$)
combinations $\psi_{S,A} =  \frac{1}{\sqrt{2}} (\Psi_x \pm \Psi_y)$ with energies 
$\pm t$($=\pm \omega/2$ since they are about to be decimated).
As in Sec.~\ref{SEC:micro}
we take these approximate eigenstates to be the lowest energy states appearing in the sum Eq.~(\ref{eq:sum})
(i.e., $n =S,A$).  
Thus, we find (including the contribution due to configurations with $\Delta<0$ between $x$ and $y$)
\begin{eqnarray}
&&\bar{G}_1(x-y;i\omega) \propto \hat{\sigma}_3 \frac{i}{\omega} \frac{1}{\Gamma^5}
\nonumber
\\
&&\qquad \times
\sum_{n=1}^{\infty} (-1)^{n+1} n^2{\rm e}^{-n^2\pi^2|x-y|/\Gamma^2},
\label{eq:g1}
\end{eqnarray}
where $\Gamma \equiv \log{\Omega_0/\omega}$ and by  \lq\lq$\propto$\rq\rq  we specifically mean equality up to 
a real and positive prefactor not captured within the RSRG approach.
We emphasize that the $\sigma_3$ structure of this contribution comes from correctly 
identifying the spinor structure of the ZE states in Eq.~(\ref{eq:spinor}) and Eq.~(\ref{eq:spinor2}).  The infinite sum
arises from the inverse  Laplace transform requried to compute $P(0,|x-y|;\omega)$.

\begin{figure}[hbt]
%
 \epsfxsize=\columnwidth
%
\centerline{\epsfbox{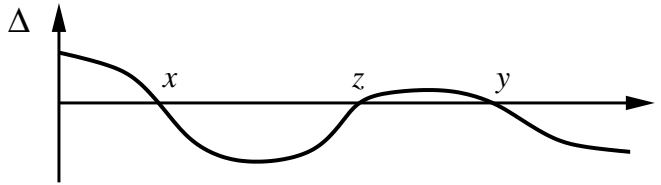}}
\vskip0.50cm \caption{Second-order class of pair potential configurations 
contributing to $\bar{G}_2(x-y;i\omega)$.  The ZE states at sites $z$ and $y$ 
are about to be decimated due to the fact that they are strongly coupled (indicated
by the relative smallness of $\Delta$ between them) but the states at sites $x$ and $z$ 
may still be weakly coupled. }
\label{fig4}
\end{figure}

Next, we turn to the evaluation of $\bar{G}_2(x-y;i\omega)$.  The type of pair-potential configuration we 
have in mind is depicted in Fig.~(\ref{fig4}).  Again, these states do not have to be 
nearest neighbor sign changes but merely nearest neighbors after having decimated until $\omega$.
The ZE states
at $x$ and $z$ are weakly coupled (by a matrix element $\delta$), owing to the relative largeness of 
$\Delta$ between them.  The ZE states at $z$ and $y$ are strongly coupled with matrix element $t= \omega/2$
and are thus about to be decimated.
Let us consider the Hamiltonian $H$ for our system 
in the subspace of these three ZE states:
\begin{equation}
H =  \begin{pmatrix}
0 &\delta & 0   \cr
\delta & 0  & t   \cr
 0  & t & 0 
\end{pmatrix}.
\end{equation}
The eigenvalues of this matrix are $0$ and $\pm \sqrt{t^2+\delta^2}$; as we are interested in
low energies,
for the purposes of the sum in Eq.~(\ref{eq:sum}) we proceed by keeping only 
the state $\psi_0$ having $0$  eigenvalue in Eq.~(\ref{eq:sum}), which has the explicit form 
\begin{equation}
\psi_0 = 
\frac{1}{\sqrt{\delta^2+t^2}} 
\begin{pmatrix}
t   \cr
0  \cr
 -\delta
\end{pmatrix}.
\label{eq:column}
\end{equation}
Of course, the elements of the column vector in Eq.~(\ref{eq:column}) denote the amplitude of each  
ZE state contribution associated with each site.  Since the ZE state spinors at $x$ and $y$ 
are each of the form of Eq.~(\ref{eq:spinor2}), the 
term appearing in the sum has the form (switching to a slightly different notation and neglecting 
normalization factors as usual)
\begin{equation}
\frac{\psi_0(x) \psi_0(y)}{i\omega}= \frac{t\delta}{\sqrt{\delta^2+t^2}}\frac{1}{i\omega} \begin{pmatrix}
1&-i   \cr
i&1  
\end{pmatrix},
\label{eq:matrix2}
\end{equation}
where we note that an extra $-$ sign in Eq.~(\ref{eq:matrix2}) came from the phase factors in Eq.~(\ref{eq:spinor})
 and Eq.~(\ref{eq:spinor2}).
  By including the possibility that $\Delta < 0$ between $z$ and $y$,
it is straighforward to see that the sum of these contributions is proportional to the unit matrix.  
Since it is about to be decimated, $t= \omega/2$. The other matrix element $\delta$ may be labeled by
$\zeta$  (as usual)  via  $\delta = \frac{\omega}{2} {\rm e}^{-\zeta}$.
Thus, we have [neglecting $\delta \ll t$ in the denominator of Eq.~(\ref{eq:matrix2})]
\begin{widetext}
\begin{equation}
\bar{G}_2(x-y;i\omega) \propto  n_{\omega} 
 \frac{\hat{\sigma}_0}{i\omega} \int_x^y dz \int_0^{\infty} d\zeta P(\zeta,x-z;\omega)P(0,z-y;\omega) ,
\label{eq:conv}
\end{equation}
\end{widetext}
where $\sigma_0$ is the unit matrix.
Taking the Laplace transform of both sides of Eq.~(\ref{eq:conv}) and evaluating the integral over $\zeta$, we find
that the Laplace transform of $\bar{G}_2(x-y;i\omega)$ is 
\begin{equation}
\bar{G}_2(\hat{y};i\omega) \propto  \frac{\sigma_0}{i\omega} \frac{1}{\Gamma^2} \frac{\hat{y}}{\sinh \Gamma\sqrt{\hat{y}}}
\frac{1}{\sinh \Gamma\sqrt{\hat{y}}  +   \sqrt{\hat{y}}\coth \Gamma\sqrt{\hat{y}}},
\label{eq:ltg2}
\end{equation}
with $\Gamma \equiv \log \Omega_0/\omega$ as before.  To obtain $\bar{G}_2(x-y;i\omega)$, we must evaluate the inverse 
Laplace transform of Eq.~(\ref{eq:ltg2}).  To do this, we use the standard technique of identifying the poles $\hat{y}_n$ of 
Eq.~(\ref{eq:ltg2}); these occur when $\sinh \Gamma\sqrt{\hat{y}_n} = 0$, i.e., $y_n = -n^2 \pi^2/\Gamma^2$ (we must
exclude the case $n=0$).  Multiplying
the associated residues by $\exp(|x-y| \hat{y}_n)$ and summing over all $n$ yields the result
\begin{equation}
\bar{G}_2(x-y;i\omega) \propto  \frac{i \sigma_0}{\omega} \frac{1}{\Gamma^5}
\sum_{n=1}^{\infty} n^2{\rm e}^{-n^2\pi^2|x-y|/\Gamma^2},
\label{eq:g2}
\end{equation}
Our final result for  $\bar{G}(x-y;i\omega)= \bar{G}_1(x-y;i\omega)+\bar{G}_2(x-y;i\omega)$
agrees with the exact results of Ref.~(\onlinecite{REF:BF}), as can be seen by taking the $M\to 0$ 
($M$ being proportional to $[\Delta(s)]_{\rm dis}$ in our notation) 
limit of Eq.~(5.30) in that paper.  In the present context, this form for $\bar{G}(x-y;i\omega)$ indicates that
the Andreev bound state Green function decays exponentially as a function of $|x-y|$ but with a 
characteristic length scale $~\log ^2 \Omega_0/\omega $ that diverges as $\omega \to 0$, indicating
the delocalization of low-energy quasiparticles along a trajectory.


\begin{thebibliography}{}
\bibitem{REF:Hosseini}
A. Hosseini, R. Harris, S. Kamal, P Dosanjh, J. Preston, R. Liang, W.N. Hardy and D. A. Bonn,
Phys. Rev. B {\bf 60}, 1349 (1999).
%
\bibitem{REF:Turner}
P.J. Turner, R. Harris, S. Kamal, M.E. Hayden, D.M. Broun, D.C. Morgan, A. Hosseini, P. Dosanjh, 
G. Mullins, J.S. Preston, R. Liang, D. A. Bonn and W.N. Hardy, cond-mat/0111353 (to appear in PRL).
%
\bibitem{REF:Hirschfeld}
P.J. Hirschfeld, W.O. Putikka and D.J. Scalapino, 
Phys. Rev. Lett. {\bf 71}, 3705 (1993); Phys. Rev. B {\bf 50}, 10250 (1994). 
%
\bibitem{REF:Durst}
A.C. Durst and P.A. Lee, Phys. Rev. B {\bf 62}, 1270 (2000).
%
\bibitem{REF:Balatsky}
A.V. Balatsky, A. Rosengren, and B.L. Altshuler,
Phys. Rev. Lett. {\bf 73\/}, 720 (1994).
\bibitem{REF:Inanc}
I. Adagideli, P.M. Goldbart, A. Shnirman and A. Yazdani,
Phys. Rev. Lett. {\bf 83\/}, 5571 (1999).
\bibitem{REF:Inanc2}
I. Adagideli, D.E. Sheehy and P.M. Goldbart,
Phys. Rev. B {\bf 66} 140512(R) (2002).
%
\bibitem{REF:Durst2}
A.C. Durst and P. A. Lee, Phys. Rev. B {\bf 65}, 094501 (2002).
%
\bibitem{REF:Yashenkin01}
A.G. Yashenkin, W. A. Atkinson, I.V. Gornyi, P.J. Hirschfeld and
 D. V. Khveshchenko,
Phys. Rev. Lett. {\bf 86\/}, 5982 (2001).
\bibitem{REF:Altland}
A. Altland, Phys. Rev. B {\bf 65}, 104525 (2002).
\bibitem{REF:Walter}
H. Walter, W. Prusseit, R. Semerad, H. Kinder, W. Assmann, H. Huber, H. Burkhardt, D. Rainer and 
J.A. Sauls,  Phys Rev. Lett. {\bf 80}, 3598 (1998).
\bibitem{REF:Sheehy}
D.E. Sheehy, I. Adagideli, P.M. Goldbart and A. Yazdani,
Phys. Rev. B {\bf 64\/}, 224518 (2001).
\bibitem{REF:Dasgupta}
C. Dasgupta and S.K. Ma, Phys. Rev. B {\bf 22}, 1305 (1980).
\bibitem{REF:Fisher}
D. S. Fisher, 
Phys. Rev. B {\bf 50}, 3799 (1994). 
%
%
\bibitem{REF:Fabrizio}
M. Fabrizio and R. M\'elin
Phys. Rev. B {\bf 56}, 5996 (1997).
%
\bibitem{REF:Steiner}
M. Steiner,  M. Fabrizio and A. O. Gogolin,
Phys. Rev. B {\bf 57}, 8290 (1998).
%
\bibitem{REF:Motrunich}
O. Motrunich, K. Damle and D.A. Huse, 
Phys. Rev. B {\bf 63}, 134424 (2001).
%
\bibitem{REF:Andreev}
A.F. Andreev,
Zh. Eksp. Teor. Fiz.~{\bf 46\/}, 1823 (1964)
[Sov. Phys. JETP~{\bf 19\/}, 1228 (1964)].
\bibitem{REF:papers}
See, e.g.,
D. Waxman and K. D. Ivanova-Moser,
Ann. Phys. {\bf 226\/}, 271 (1993);
D.G. Shelton and A.M. Tsvelik,
Phys. Rev. B {\bf 57}, 14242 (1998);
%
A.J. Millis and H. Monien
Phys. Rev. B {\bf 61}, 12496 (2000);
L. Bartosch,
Ann. Phys. (Leipzig) {\bf 10\/}, 799 (2001).
\bibitem{REF:OE}
A.A. Ovchinnikov and N.S. \'Erikhman
Zh. Eksp. Teor. Fiz.~{\bf 73\/}, 650 (1977)
[Sov. Phys. JETP~{\bf 46\/}, 340 (1977)].
%
%
\bibitem{REF:BF}
L. Balents and M.P.A. Fisher,
Phys. Rev. B~{\bf 56\/}, 12970 (1997).
\bibitem{REF:SvH82}
P. Salomonson and J.W. Van Holten,
Nucl. Phys. {\bf B 196}, 509 (1982).
\bibitem{REF:Witten}
E. Witten,
Nucl. Phys. {\bf B 188}, 513 (1981).
\bibitem{note:zero}
One can envision a situation in which $\Delta(s)$ vanishes without changing sign.  In an infinite system with exactly one
such zero of $\Delta$, the state Eq.~(\ref{eq:susystate}) is not normalizable.  Thus, we shall ignore such zeroes of 
$\Delta$; alternately, one can think of such zeroes as a pair of sign changes in which the distance between the 
successive sign changes is infinitesimally small.  The tunneling matrix element between them would correspondingly be
very large, so that they would be immediately decimated.
\bibitem{REF:Mahan} See, e.g.,
G.D.  Mahan, {\it Many Particle Physics\/} 
(Plenum, New York, 1990).
\end{thebibliography}
\end{document}